\documentclass{article}

\usepackage{amssymb,amsfonts,amsmath}
\usepackage{cite,enumerate,float}
\usepackage{color}
\usepackage{tikz}
\usetikzlibrary{arrows,snakes,backgrounds}

\def\be{\begin{eqnarray}}
\def\ee{\end{eqnarray}}
\def\nn{\nonumber}

\definecolor{red}{rgb}{1,0,0}
\definecolor{orange}{rgb}{1,0.5,0}
\definecolor{violet}{rgb}{0.7,0,1}

\hoffset= - 1.0in         

\begin{document}

\title{\vspace{1.5cm}\bf
Commutative subalgebras from Serre relations
}

\author{
A. Mironov$^{b,c,d,}$\footnote{mironov@lpi.ru,mironov@itep.ru},
V. Mishnyakov$^{a,b,c,e,}$\footnote{mishnyakovvv@gmail.com},
A. Morozov$^{a,c,d,}$\footnote{morozov@itep.ru},
A. Popolitov$^{a,c,d,}$\footnote{popolit@gmail.com}
}

\date{ }

\maketitle

\vspace{-6.5cm}

\begin{center}
\hfill FIAN/TD-11/23\\
\hfill IITP/TH-12/23\\
\hfill ITEP/TH-16/23\\
\hfill MIPT/TH-13/23
\end{center}

\vspace{4.5cm}

\begin{center}
$^a$ {\small {\it MIPT, Dolgoprudny, 141701, Russia}}\\
$^b$ {\small {\it Lebedev Physics Institute, Moscow 119991, Russia}}\\
$^c$ {\small {\it NRC ``Kurchatov Institute", 123182, Moscow, Russia}}\\
$^d$ {\small {\it Institute for Information Transmission Problems, Moscow 127994, Russia}}\\
$^e$ {\small{\it Institute for Theoretical and Mathematical Physics, Lomonosov Moscow State University, Moscow 119991, Russia}}
\end{center}

\vspace{.1cm}

\begin{abstract}
We demonstrate that commutativity of numerous one-dimensional subalgebras in $W_{1+\infty}$ algebra,
i.e. the existence of many non-trivial integrable systems
described in recent arXiv:2303.05273
follows from the subset of relations in algebra
known as Serre relations.
No other relations are needed for commutativity. The Serre relations survive the deformation to the affine Yangian $Y(\hat{\mathfrak{gl}}_1)$, hence the commutative subalgebras do as well.
A special case of the Yangian parameters corresponds to the $\beta$-deformation. The preservation of Serre relations can be  thought of a selection rule for proper systems of commuting $\beta$-deformed Hamiltonians.
On the contrary, commutativity in the extended family
associated with ``rational (non-integer) rays" is {\it not} reduced to the Serre relations,
and uses also other relations in the $W_{1+\infty}$ algebra.
Thus their $\beta$-deformation is less straightforward.
\end{abstract}

\bigskip

\section{Introduction
\label{intro}}

After a remarkable suggestion of \cite{WLZZ} to use $W$-representations \cite{Wrep}
for production of numerous superintegrable \cite{MMsi} matrix models \cite{China},
the new discovery of \cite{MM,MMMP} was that they are associated with a vast variety
of new integrable systems, which are the far-going generalizations of the rational Calogero model \cite{Cal}.
The fact that commutativity is observed in all these systems in various realizations \cite{MMMP}:
in the $w_\infty$ contraction, in the one-body representation, in the second quantization (time variables) representation, in eigenvalues
and even in matrices (after restriction to the subspace of invariant functions)
assumes a claim that it can have its origin in the basic algebraic properties
of the underlying $W_{1+\infty}$ algebra, while its further $\beta$-deformation with the affine Yangian of $\mathfrak{gl}_1$.
The suggestion of \cite{MMMP} was to derive commutativities from a set of the
Serre relations, the double-commutator relations within the Borel subalgebra
independent of all the  deformations (parameters $\sigma_3 \sim \beta-1$ and $\sigma_2$ in the Yangian $\beta$-deformation).
Besides a conceptual beauty, this suggestion could provide a selection rule for the
$\beta$-deformed Hamiltonians, which preserves their commutativity.
In \cite{MMMP}, this suggestion was
made for ``integer rays", but
supported just by a couple of oversimplified examples,
while their generalization looked quite difficult.
In this paper, we provide a proof that
{\bf the integer-ray 1-dimensional sets of commutative Hamiltonians exist due to ``the Serre relations"
and they have a distinguished $\beta$-deformation preserving commutativity}.
This is demonstrated in Section~\ref{sec:proof-integer-rays}.

At the same time, in Section~\ref{sec:attemps-rational-rays} we confirm the conclusion of  \cite{MMMP} that for ``rational rays" the Serre relations
are not enough, commutativity in this sector makes use of the main quadratic
commutation relation,
which involves anticommutator and is $\beta$-dependent.
Thus, the $\beta$-deformation breaks commutativity of the naively constructed families, as it was already observed
in \cite{MMMP} in particular representations.

\section{$W_{1+\infty}$ algebra and affine Yangian of $\mathfrak{gl}_1$}

The $W_{1+\infty}$-algebra is the Lie algebra which, as is typical for many Lie algebras, can be described in two different ways: by commutators of all elements of a linear basis in the algebra, $O_{n,m}$ \cite{Pope,FKN2,BK,BKK,KR1,FKRN,Awata,KR2,Miki}, or
 \cite{SV,AS,Tsim,Prochazka} by commutators of a few generating elements $\Psi_i$, $F_i$, $E_i$, $i\in\mathbb{Z}_{\ge 0}$:
\be
\phantom{.}[\hat\Psi_j,\hat\Psi_k]&=&0\nn\\
\phantom{.}[\hat E_j,\hat F_k]&=&\hat \Psi_{j+k}\nn\\
\phantom{.}[\hat\Psi_{0},\hat E_j]&=&0,\ \ \ \ \ [\hat\Psi_{0},\hat F_j]=0\nn\\
\phantom{.}[\hat\Psi_{1},\hat E_j]&=&0,\ \ \ \ \ [\hat\Psi_{1},\hat F_j]=0\nn\\
\phantom{.}[\hat\Psi_2,\hat E_j]&=&2 \hat E_j,\ \ \ \ \ [\hat\Psi_2,\hat F_j]=-2\hat F_j
\ee
with additional relations on them: quadratic
\be\label{YW}
\phantom{.}[\hat E_{j+3},\hat E_k]-3[\hat E_{j+2},\hat E_{k+1}]+3[\hat E_{j+1},\hat E_{k+2}]-[\hat E_{j},\hat E_{k+3}]
-[\hat E_{j+1},\hat E_{k}]+[\hat E_{j},\hat E_{k+1}]&=&0\nn\\
\phantom{.}[\hat F_{j+3},\hat F_k]-3[\hat F_{j+2},\hat F_{k+1}]+3[\hat F_{j+1},\hat F_{k+2}]-[\hat F_{j},\hat F_{k+3}]
-[\hat F_{j+1},\hat F_{k}]+[\hat F_{j},\hat F_{k+1}]&=&0\nn\\
\phantom{.}[\hat \Psi_{j+3},\hat E_k]-3[\hat \Psi_{j+2},\hat E_{k+1}]+3[\hat \Psi_{j+1},\hat E_{k+2}]-[\hat \Psi_{j},\hat E_{k+3}]
-[\hat \Psi_{j+1},\hat E_{k}]+[\hat \Psi_{j},\hat E_{k+1}]&=&0\nn\\
\phantom{.}[\hat \Psi_{j+3},\hat F_k]-3[\hat \Psi_{j+2},\hat F_{k+1}]+3[\hat \Psi_{j+1},\hat F_{k+2}]-[\hat \Psi_{j},\hat F_{k+3}]
-[\hat \Psi_{j+1},\hat F_{k}]+[\hat \Psi_{j},\hat F_{k+1}]&=&0
\label{eq:serre-quadratic}
\ee
and cubic (the Serre relations)
\be
S_{ijk}:= \!\!\!\!\!\!\!\!\!\!\!\!\!\!\! \!\!\!\!\!\!\!\!\!\!\!\!\!\!\! \!\!\!\!\!\!\!\!\!\!\!\!\!\!\!
 \!\!\!\!\!\!\!\!\!\!\!\!\!\!\! \!\!\!\!\!\!\!\!
\underbrace{\hbox{Sym}_{i,j,k}[\hat E_i,[\hat E_j,\hat E_{k+1}]]}_
{[\hat E_i,[\hat E_j,\hat E_{k+1}]]+ [\hat E_i,[\hat E_k,\hat E_{j+1}]]+[\hat E_j,[\hat E_i,\hat E_{k+1}]]
+ [\hat E_j,[\hat E_k,\hat E_{i+1}]]+ [\hat E_k,[\hat E_i,\hat E_{j+1}]]+[\hat E_k,[\hat E_j,\hat E_{i+1}]]}
\!\!\!\!\!\!\!\!\!\!\!\!\!\!\! \!\!\!\!\!\!\!\!\!\!\!\!\!\!\! \!\!\!\!\!\!\!\!\!\!\!\!\!\!\!
 \!\!\!\!\!\!\!\!\!\!\!\!\!\!\! \!\!\!\!\!\!\!\!
 = 0
\nn\\
\hbox{Sym}_{i,j,k}[\hat F_i,[\hat F_j,\hat F_{k+1}]]=0
\label{eq:serre-cubic}
\ee
where the symbol $\hbox{Sym}_{i,j,k}$ means the symmetrization over the three indices $i,j,k$. Note that, since $\hat\Psi_0$ commutes with all elements of algebra, it is just the central charge of the algebra:
\be
\hat \Psi_0=c
\ee
In fact, the whole algebra is generated by three operators $\hat \Psi_3$, $\hat E_0$ and $\hat F_0$, \cite{SV,AS}\footnote{Note that $\hat \Psi_3=6\hat W_0$ in \cite{MMMP}, while $\hat E_i$ and $\hat F_i$ coincide with those in \cite{MMMP}.},
of which the last two can be considered as ``simple roots", and all the other \eqref{eq:serre-quadratic}-\eqref{eq:serre-cubic}
play the role of the Serre relations for  ``the roots" recursively defined as
\be\label{rdef}
\hat E_{k+1} = \frac{1}{6}[\hat \Psi_3,\hat E_k]  \nn \\
\hat F_{k+1} = -\frac{1}{6}[\hat \Psi_3,\hat F_k]
\ee
However, in this text, we reserve the term ``Serre relations" for \eqref{eq:serre-cubic} only.

The second representation of the $W_{1+\infty}$ algebra has an advantage of simple deformation to the affine Yangian of $\mathfrak{gl}_1$ (it is also isomorphic to the algebra $SH^c$ of \cite{SV}). This latter is defined \cite{SV,AS,Tsim,Prochazka} to be an associative algebra with quadratic relations
\be\label{Y}
\phantom{.}[\hat E_{j+3},\hat E_k]-3[\hat E_{j+2},\hat E_{k+1}]+3[\hat E_{j+1},\hat E_{k+2}]-[\hat E_{j},\hat E_{k+3}]
-[\hat E_{j+1},\hat E_{k}]+[\hat E_{j},\hat E_{k+1}]-\nn\\
-\beta(\beta-1)\left(\{\hat E_j,\hat E_k\}+[\hat E_{j+1},\hat E_{k}]-[\hat E_{j},\hat E_{k+1}]\right)&=&0\nn\\
\phantom{.}[\hat F_{j+3},\hat F_k]-3[\hat F_{j+2},\hat F_{k+1}]+3[\hat F_{j+1},\hat F_{k+2}]-[\hat F_{j},\hat F_{k+3}]
-[\hat F_{j+1},\hat F_{k}]+[\hat F_{j},\hat F_{k+1}]-\nn\\
-\beta(\beta-1)\left(\{\hat F_j,\hat F_k\}+[\hat F_{j+1},\hat F_{k}]-[\hat F_{j},\hat F_{k+1}]\right)&=&0\nn\\
\phantom{.}[\hat \Psi_{j+3},\hat E_k]-3[\hat \Psi_{j+2},\hat E_{k+1}]+3[\hat \Psi_{j+1},\hat E_{k+2}]-[\hat \Psi_{j},\hat E_{k+3}]
-[\hat \Psi_{j+1},\hat E_{k}]+[\hat \Psi_{j},\hat E_{k+1}]-\nn\\
-\beta(\beta-1)\left(\{\hat \Psi_j,\hat E_k\}+[\hat \Psi_{j+1},\hat E_{k}]-[\hat \Psi_{j},\hat E_{k+1}]\right)&=&0\nn\\
\phantom{.}[\hat \Psi_{j+3},\hat F_k]-3[\hat \Psi_{j+2},\hat F_{k+1}]+3[\hat \Psi_{j+1},\hat F_{k+2}]-[\hat \Psi_{j},\hat F_{k+3}]
-[\hat \Psi_{j+1},\hat F_{k}]+[\hat \Psi_{j},\hat F_{k+1}]-\nn\\
-\beta(\beta-1)\left(\{\hat \Psi_j,\hat F_k\}+[\hat \Psi_{j+1},\hat F_{k}]-[\hat \Psi_{j},\hat F_{k+1}]\right)&=&0
\ee
instead of (\ref{YW}), while all other relations do not change. Here $\{\ldots\}$ denotes the anticommutator, and $\beta$ is some deformation constant. In fact, the most general deformation depends on two parameters $\sigma_2$ and $\sigma_3$ \cite{Tsim,Prochazka}, which are related with $\beta$ by\footnote{In another parametrization, $\sigma_1=h_1+h_2+h_3=0$, $\sigma_2=h_1h_2+h_1h_3+h_2h_3$, $\sigma_3=h_1h_2h_3$, the relation is $h_1=1$, $h_2=-\beta$, $h_3=\beta-1$.} $\sigma_2=-1-\beta(\beta-1)$, $\sigma_3=-\beta(\beta-1)$.

One can again generate the whole algebra \cite{SV,AS} starting from the three generating elements $\hat \Psi_3$, $\hat E_0$ and $\hat F_0$, \cite{SV,AS}\footnote{In the $\beta$-deformed case, the identification with \cite{MMMP} is as follows: $\hat \Psi_3-\beta(\beta-1)\hat \Psi_2=6\hat W_0$, while $\hat E_i$ and $\hat F_i$ still coincide with those in \cite{MMMP}.}, however,
formulas (\ref{rdef}) are deformed due to (\ref{Y}):
\be\label{rdefY}
\hat E_{k+1} = \frac{1}{6}[\hat \Psi_3,\hat E_k]-{\frac{c}{3}}\beta(\beta-1)\hat E_k  \nn \\
\hat F_{k+1} = -\frac{1}{6}[\hat \Psi_3,\hat F_k]+{\frac{c}{3}}\beta(\beta-1)\hat F_k
\ee

The cubic relations are distinguished in several respects, this is why prefer to use the term ``Serre relations" only for them:

(a) they contain double commutators and hence are \textit{cubic} in generating elements $E_i$, $F_i$;

(b) they do not depend on the deformation parameter $\beta$ (or on the parameters $\sigma_3$ and $\sigma_2$ in the generic case);

The third essential property, which also holds for \eqref{eq:serre-quadratic}, is

(c) the shift invariance: $\hbox{Sym}_{m+i,m+j,m+k}=0$ holds along with $\hbox{Sym}_{i,j,k}$ for any $m$.
This means that any $i,j,k$-dependent symbolic corollary of $\hbox{Sym}_{i,j,k}=0$ is automatically true
for $m+i,m+j,m+k$.

\section{Commutative subalgebras}

What we prove in this paper (see Section~\ref{sec:proof-integer-rays}) is that
relations \eqref{eq:serre-cubic} imply commutativity of certain iterated commutators,
which can be identified with sets 
of commutative Hamiltonians $ H^{(m)}_k$ from \cite{MMMP} (every $m$ labels an ``integer ray" family of Hamiltonians).
This means that these Hamiltonians are commutative in the $\beta$-deformed case as well.

In fact, due to property (c), we only need to show this, say for,
\be
{ H}^{(0)}_k:={\rm ad}_{\hat E_{1}}^{k-1}\hat E_0
\label{calHk}
\ee
then commutativity of $H^{(m)}_k={\rm ad}_{\hat E_{m+1}}^{k-1}\hat E_m$ automatically follows. The commutative
family at the given $m$: $H^{(m)}_k$ was called the integer ray in \cite{MMMP}.
Note that, in the time variable representation \cite{MMMP}, ${ H}^{(0)}_k=p_{k}$
i.e. commutativity of (\ref{calHk}) becomes the commutativity of time variables $p_k$.
In other words, {\it if one knew that the commutativity {\it is} a corollary of
  \eqref{eq:serre-cubic} only},
then it would be a direct corollary of (\ref{calHk}) which is {\it closely related}
to commutativity of times.

Likewise one can consider a family of Hamiltonians ${H}^{(p,q)}_k$ labeled by coprime $(p,q)$ \cite{MMMP} (which is called ``rational ray"). For instance, for the family ${H}^{(2m+1,2)}_k$, it would have been enough to consider the case of
\be
H^{(1,2)}_k={\rm ad}_{\hat E_2^{(2)}}^{k-1}\hat E_1^{(2)}
\ee
where \cite{MMMP}
\be
\hat E_0^{(2)}&=&\phantom{1\over 6}\ [\hat E_1,\hat E_0]\nn\\
\hat E_1^{(2)}&=&{1\over 6}\ [\hat \Psi_3,\hat E_0^{(2)}]\nn\\
\hat E_2^{(2)}&=&{1\over 6}\ [\hat \Psi_3,\hat E_1^{(2)}]
\ee
so that, using the Jacobi identity,
\be
\hat E_1^{(2)}={1\over 6}\Big(\hat \Psi_3\hat E_1\hat E_0\Big)={1\over 6}\left(\Big(\hat E_0
\underbrace{\hat E_1\hat \Psi_3}_{-\hat E_2}\Big)+\Big(\hat E_1\underbrace{\hat \Psi_3\hat E_0}_{\hat E_1}\Big)\right)=
[\hat E_2,\hat E_0]\nn\\
\hat E_2^{(2)}={1\over 6}\Big(\hat\Psi_3\hat E_2\hat E_0\Big)={1\over 6}\left(\Big(\hat E_0
\underbrace{\hat E_2\hat \Psi_3}_{-\hat E_3}+\Big(\hat E_2\underbrace{\hat\Psi_3\hat E_0}_{\hat E_1}\Big)\right)=
[\hat E_2,\hat E_1]+[\hat E_3,\hat E_0]
\ee
and, hence,
\be
H^{(1,2)}_k={\rm ad}_{[\hat E_{3},\hat E_0]+[\hat E_2,\hat E_1]}^{k-1}[\hat E_2,\hat E_0]
\label{calH2}
\ee
However, as we demonstrate in sec.\ref{sec:attempts-rational-rays},
the Serre relations \eqref{eq:serre-cubic} are not enough to prove the
commutativity of this ray Hamiltonians, and one also needs to add \eqref{eq:serre-quadratic},
which non-trivially deforms as one switches $\beta$-deformation on.
Hence, at $\beta \neq 1$, the operators on this ray become non-commutative.

\bigskip

A drastic simplification that we use throughout the paper is that, due to the Jacobi identities,
it is enough to show the commutativity
\be
[{H}^{(0)}_k,{H}^{(0)}_{k+1}] = 0
\label{calH1H1}
\ee
and then all other   $[{H}^{(0)}_{k_1},{H}^{(0)}_{k_2}] = 0$ with $k_1+k_2=2k+1$ follow
from the Jacobi identities and from the commutativity at the previous level $2k$.
Moreover, at even $k_1+k_2=2k$ there is nothing new to check:
the Jacobi identities reduce all commutativities to those at the previous odd level $2k-1$.

In what follows, we denote the repeated commutator as
$$(n_1,n_2,n_3, \ldots, n_{s-1},n_s):=
\left[\hat E_{n_1},\big[\hat E_{n_2},[\hat E_{n_3},  \ldots [\hat E_{n_{s-1}},\hat E_{n_s}]\ldots]\big]\right],
$$
e.g.
${H}^{(0)}_k = {\rm ad}_{\hat E_{1}}^{k-1}\hat E_0 = (\underbrace{1,\ldots,1}_{k-1}, 0)$,
and
$[{H}^{(0)}_1,{H}^{(0)}_k] =  (0,\underbrace{1,\ldots,1}_{k-1}, 0)$.
To succeed in proving the commutativity, we explain that the latter quantity vanishes due to the Serre identities,
$(0, 1,\ldots,1 , 0)\stackrel{S}{=}0$.

\section{Proof  of commutativity for all integer rays
  \label{sec:proof-integer-rays}
} 

\subsection{Preliminary examples}

\begin{itemize}

\item{}
The first relation at level {\bf three},
\be
\boxed{
[H_1^{(m)},H_2^{(m)}] := [E_m, [E_{m+1},E_m]] := (m,m+1,m) \stackrel{S}{=} 0
\label{H12Serr}
}
\ee
is just (minus) the Serre identity $S^{(m)}_{000}:=[E_m,[E_m,E_{m+1}]]=0$,
which we abbreviate to $S^{(m)}_{000}:=(m,m,m+1)=-(m,m+1,m) = 0$.
Since $m$ enters trivially in this and subsequent formulas, we omit this
uniform shift of indices by $m$ in what follows.
In particular, hereafter, (\ref{H12Serr}) is abbreviated to $S_{000} =-(010)\stackrel{S}{=}0$.

\item{}
The {\bf fourth} level is even, and the commutativity relation trivializes:
\be
\boxed{
[H_1^{(m)},H_3^{(m)}] \ \stackrel{J+(\ref{H12Serr})}{=}\  [H_2^{(m)},H_2^{(m)}] = 0
\label{H13Serr}
}
\ee

\item{}
The non-trivial result at the {\bf fifth} level is that
\be
(01110) \stackrel{J}{=} \frac{4\cdot(02S_{000})-3\cdot (11S_{000}) -3\cdot(20S_{000})
+9\cdot (10S_{001})-15\cdot (01S_{001})+3(00S_{002})}{30} \stackrel{s}{=} 0
\label{01110}
\ee
i.e. $(01110) \stackrel{J+S}{=} 0 \ \ \Longrightarrow \ \ (m,m+1,m+1,m+1,m) \stackrel{J+S}{=} 0$,
so that
$$
[H^{(m)}_1,H^{(m)}_4] := \left[E_m,[E_{m+1},[E_{m+1},[E_{m+1},E_m]]]\right] :=
(m,m+1,m+1,m+1,m) \stackrel{J}= \frac{4}{30}\cdot [E_{m},[E_{m+2},S^{(m)}_{000}]] + \ldots
$$
where dots stand for the other terms at the r.h.s. of (\ref{01110}),
and
\be
\boxed{
[H^{(m)}_2,H^{(m)}_3]\ \stackrel{J+(\ref{H13Serr})}{=}\  [H^{(m)}_1,H^{(m)}_4] = (m,m+1,m+1,m+1,m) \stackrel{S}{=} 0
}
\label{H14Serr}
\ee

\item{}
The {\bf sixth} level is even, and it is again trivial:
\be
\boxed{
[H^{(m)}_1,H^{(m)}_5]\ \stackrel{J+(\ref{H14Serr})}{=}\  [H^{(m)}_2,H^{(m)}_4]
\ \stackrel{J+(\ref{H14Serr})}{=}\  [H^{(m)}_3,H^{(m)}_3] = 0
}
\label{H15Serr}
\ee
though the explicit (i.e. not involving recursive arguments)
check is rather lengthy, namely
{\footnotesize
\be
[H_1,H_5] = (011110) \stackrel{J}{=} \frac{1}{6}(210S_{000})-\frac{1}{6}(201S_{000}) - \frac{1}{3}\cdot(120S_{000})
+\frac{1}{6}(102S_{000}) +\frac{1}{2}\cdot(012S_{000}) - \frac{1}{6}(021S_{000})
+ \nn \\
+(200S_{001})-2\cdot(020S_{001})+(002S_{001})
-  \frac{1}{2}(011S_{001})
+\frac{1}{2}(100S_{011}) - \frac{1}{2}(001S_{011})   +\frac{1}{6}(000S_{111}) \ \stackrel{S}{=} 0
\nn
\ee
}
where, for the sake of brevity, we omitted the superscript $m$.

This expression in terms of Serre relation corollaries is actually not unique: only 13 out of 17 possible structures are present at the r.h.s. in this version,
and there is no clear way how to prefer one decomposition over another.

\item{}
The {\bf seventh} level is odd, and it is again non-trivial:\footnote{
A possible solution (one of many) is

\vspace{-0.5cm}
\begin{align} \scriptscriptstyle
  u_{0, 0, 0, 2, 0, 1, 1} & \ \scriptscriptstyle = -1/35, u_{0, 0, 1, 0, 1, 1, 1} = 1/28, u_{0, 0, 1, 1, 0, 1, 1} = 8/35, u_{0, 0, 1, 2, 0, 0, 1} = -1/5, u_{0, 0, 2, 1, 0, 0, 1} = 6/35, u_{0, 0, 2, 2, 0, 0, 0} = 4/105, u_{0, 1, 0, 1, 0, 1, 1} = -31/35,
 \notag \\ \notag \scriptscriptstyle
  u_{0, 1, 0, 2, 0, 0, 1} & \ \scriptscriptstyle = 1/7, u_{0, 1, 1, 0, 0, 1, 1} = 23/35, u_{0, 1, 1, 1, 0, 0, 1} = -1/2, u_{0, 1, 1, 2, 0, 0, 0} = 79/420, u_{0, 1, 2, 0, 0, 0, 1} = 8/35, u_{0, 1, 2, 1, 0, 0, 0} = -22/105, u_{0, 2, 0, 1, 0, 0, 1} = 12/35,
\\ \notag \scriptscriptstyle
u_{0, 2, 0, 2, 0, 0, 0} = & \ \scriptscriptstyle -4/21, u_{0, 2, 1, 0, 0, 0, 1} = -16/35, u_{0, 2, 1, 1, 0, 0, 0} = 47/420, u_{0, 2, 2, 0, 0, 0, 0} = 3/35, u_{1, 0, 1, 0, 0, 1, 1} = 3/14, u_{1, 0, 1, 1, 0, 0, 1} = 109/140,
\\ \notag \scriptscriptstyle
u_{1, 0, 1, 2, 0, 0, 0} = & \ \scriptscriptstyle -17/105, u_{1, 0, 2, 0, 0, 0, 1} = -6/35, u_{1, 0, 2, 1, 0, 0, 0} = 11/60, u_{1, 1, 0, 1, 0, 0, 1} = -109/70, u_{1, 1, 0, 2, 0, 0, 0} = 47/210, u_{1, 1, 1, 0, 0, 0, 1} = 26/35,
\\ \notag \scriptscriptstyle
u_{1, 1, 1, 1, 0, 0, 0} = & \ \scriptscriptstyle 1/84,
u_{1, 2, 0, 1, 0, 0, 0} =  -1/105, u_{1, 2, 1, 0, 0, 0, 0} = 0, u_{2, 0, 0, 1, 0, 0, 1} = -6/35, u_{2, 0, 0, 2, 0, 0, 0} = 4/35, u_{2, 0, 1, 1, 0, 0, 0} = -13/105, u_{2, 0, 2, 0, 0, 0, 0} = -2/35,
\\ \notag \scriptscriptstyle
u_{2, 1, 0, 0, 0, 0, 1} = & \ \scriptscriptstyle 6/35, u_{2, 1, 0, 1, 0, 0, 0} = 8/105, u_{2, 1, 1, 0, 0, 0, 0} = -1/12
\end{align}

\noindent All other $u=0$. In fact, all but $u_{2, 2, 0, 0, 0, 0, 0}$ are free parameters (which can be put equal to something else).
}

\vspace{-0.5cm}
\begin{align}
  [H_1,H_6] = (0111110) \stackrel{J}{=} \sum_{i,j,k,l=0}^2 \sum_{a\leq b\leq c=0}^1
  \delta_{i+j+k+l+a+b+c, 4}
  \cdot u_{i,j,k,l,a,b,c} \cdot (ijkl S_{abc})
\end{align}

\end{itemize}

The main lesson from these examples is that, in order to prove commutativity of the Hamiltonians for {\it all} rational rays,
one needs to prove that
\be
\boxed{
(01^k0) = 0
}
\label{intcom}
\ee
for all $k$, and this will be the task of the next subsection.

\subsection{Proof}

We will now provide a complete {\bf proof} of the identity $(\ldots \underline{01^k0})=0$ by induction in $k$.
Actually we will prove it together with another identity:
\be
\left\{ \begin{array}{lcl}
(\ldots \underline{01^k0})&=&0 \\
(\ldots \underline{1^{k+1}0}) &=& \frac{k(k+1)}{2}\cdot (\ldots 01^{k-1}02)
\end{array}\right.
\label{indhyp}
\ee

\begin{itemize}

\item{}
The starting point is $k=1$:
\be
\left\{\begin{array}{lcl}
(\ldots \underline{010}) &\stackrel{S_{000}}{=}& 0 \\
(\ldots \underline{110}) &\stackrel{S_{001}}{=}& (\ldots 002)
\end{array}\right.
\label{k=1}
\ee
Both identities in this case are just the Serre relations.

Formally, in order to complete the proof, one can proceed directly to the last item in this subsection marked with two bullets.
However, we prefer to proceed slower, and present more details on the way.

\item{}
At $k=2$, we have:
\be
\left\{ \begin{array}{cclccc}
(\ldots \underline{\underline{0110}}) &=& (\ldots \!\!\!\!\underbrace{[0,1]10}_{-\left[[0,1],[0,1]\right]=0}\!\!\!\!)
+ (\ldots 1\underline{010})
&\stackrel{S_{001}}{=}& 0
\\
(\ldots \underline{\underline{1110}})  &=& ? &=& 3\cdot(\ldots 0102)
\end{array}\right.
\label{k=2}
\ee
We double underlined the structure of our interest at this level $k=2$,
while those which are known at the previous step $k=1$,
are underlined only once.
In the  first line, we can either directly apply the Serre relation $(001)=-(010)=0$ to the second term,
or recognize it is an induction hypothesis at $k=1$, thus reducing the case $k=2$  to $k=1$.
But what should we do with the second line?
How do we get the desired answer, {\it announced} after the question mark?

Here we need an additional calculation, which will be generally used for making an induction step $k\longrightarrow k+1$.
Note that all our manipulations are done with the rightmost entries $B$ of $(\ldots B)$, which means that it is sufficient to work merely with identities for $(B)$: the left, dot part of the sequence $(\ldots B)$ means just inserting these identities for $(B)$ into further commutators. Hence, from now on, we omit these inessential leftmost dots. Apply
\be
(A0B)=[A,[0,B]]\stackrel{J}{=}[B,[0,A]]-[0,[B,A]] = (B0A)-(0BA)
\ee
to $B_k=(\underbrace{(1\ldots 1}_k0) :=(1^k0)$:
\vspace{-1.0cm}
\be
(A01^k0) \stackrel{J}{=} ((1^k0)0A)-(0(1^k0)A)
\label{Arel}
\ee
If we substitute $A=2$ and $k=1$, then
\be
(2\underline{010}) = ([1, 0]0 2)-(0[1,0]2)
= (1002)-2\cdot (0102) + (0012)
  \stackrel{S}{=}
 (\underline{\underline{1110}})-3\cdot ( 0102)
\ee
Since the l.h.s. is vanishing due to the Serre relation in the first line of (\ref{k=1}),
we get the desired identity for the second line of (\ref{k=2}):
\be
(\underline{\underline{1110}})=3\cdot ( 0102)
\label{k=22}
\ee
This completes the first step $k=1 \ \longrightarrow \ k=2$ of the induction.

\item{}
At $k=3$ we need
\be
\left\{ \begin{array}{cclcccl}
(\underline{\underline{ 01110}}) &\stackrel{(\ref{k=22})}{=} &
3\cdot (00102) &=& ? &=& 0
\\
(\underline{\underline{11110}})  &\stackrel{(\ref{k=22})}{=} &
3\cdot (10102) &=& ? &=& 6\cdot (01102)
\end{array}\right.
\label{k=3}
\ee
In order to get the desired answers, written after the question mark,
one needs to use (\ref{Arel}) in two different ways.
First, substitute $A=[0,2]$ and $k=1$:
\be
([0,2]\underline{010}) = ([1,0]002)-(0[1,0]02)
= (10002)-2\cdot (01002) + (00102)
 \stackrel{S}{=}
(1\underline{0110})-2\cdot \underline{01110})+(00102)
\nn
\ee
Since vanishing of all the underlined quantities is already proved,
one gets $(00102)=0$, thus the first line in (\ref{k=3}) is also zero.

For the second line, one needs $A=2$ and $k=2$ in (\ref{Arel}):
\be
\begin{array}{ccccccccc}
 (2\underline{0110}) &=& ([1,[1, 0]]0 2)-(0[1,[1,0]]2)
&=& (11\underbrace{002})&-2\cdot (10102)  &+2\cdot(01\underbrace{012}) &-(00\underbrace{112})&
  \stackrel{S}{=}   \\
&&&&\!\!\!\!\!\!\!\! S_{001}\ \downarrow &{\rm id} \ \downarrow
&\ \  S_{011} \ \downarrow &\!\!\!S_{111}\ \downarrow  \\
&&&\stackrel{S}{=}& (\underline{\underline{11110}})&- 2\cdot (10102)&  -2\cdot(01102) &\ \ \ - 0
\end{array}
\nn
\ee
Comparing with the second line in (\ref{k=3}) and eliminating $(10102)$,
one gets the desired result
\be
(\underline{\underline{11110}}) = 6\cdot (01102)
\label{k=33}
\ee

\item{}
At $k=4$, we want
\be
\left\{ \begin{array}{cclcccl}
(\underline{\underline{011110}}) &\stackrel{(\ref{k=3})}{=} &
6\cdot (001102) &=& ? &=& 0
\\
(\underline{\underline{111110}})  &\stackrel{(\ref{k=3})}{=} &
6\cdot (101102) &=& ? &=& 10\cdot (011102)
\end{array}\right.
\label{k=4}
\ee
In order to derive this, we make two substitutions into  (\ref{Arel}):
\be
\left\{
\begin{array}{ccc|cccccccccc}
A=[0,2]& k=2 & B=(110) & \ \ & ( [0,2] \underline{0 110 }) &=& ((110)002)-(0(110)02)
&=&
\\ &&& \\ \hline &&& \\
A=2 & k=3 & B=(1110) & &  (2\underline{01110}) &=& ((1110)02)-(0(1110)2)
&=&
\end{array}
\right.
\nn
\ee
{\footnotesize
\be
\left\{
\begin{array}{cccccccccc}
& (11\underline{0110}) &  -2\cdot (1\underline{01110})  && \!  0   && +\frac{2}{3}\cdot(\underline{\underline{011110}})
&  -\frac{1}{6}\cdot (\underline{011110})      \\
&\!\! S_{001}  \uparrow & \ \ \ \  S_{001}  \uparrow &&\!\!\!\!\!\!\!\!\! {\rm id} \ \uparrow
&&\ \ \  (\ref{k=22}) \uparrow & \!\!\!\!(\ref{k=33}) \uparrow  \\
= &(110\overbrace{002})&-2\cdot (101\overbrace{002}) &&\overbrace{+ (011002) - (011002)}&& +2\cdot(01\overbrace{0102})
&- (0\overbrace{01102})
\\ &&& \\ \hline &&& \\
= &(111\underbrace{002})&-3\cdot(11\underbrace{0102})&+3\cdot(1\underbrace{01102})&-(011102) - (011102)
&+3\cdot(011\underbrace{012}) &- 3\cdot(010\underbrace{112}) &+ (001\underbrace{112})  \\
& \!\!\! S_{001} \downarrow  & \ \ \ (\ref{k=22}) \downarrow &\ \ (\ref{k=33}) \downarrow
&{\rm id}\ \downarrow     &\ \ \ \ S_{011}\downarrow &\ \ \ \ S_{111} \downarrow & S_{111}\downarrow  \\
& (\underline{\underline{111110}})    &- (\underline{\underline{111110}})
& +\frac{1}{2}\cdot (\underline{\underline{111110}})  &  -2\cdot(011102) &-3\cdot(011102)
&\ \ \ \  \ \ \ \ \ \ \ 0 & \ \ \ \ \ \ 0
\end{array}
\right.
\nn
\ee
}

Since the single-underlined quantities are vanishing, we obtain for the double-underlined ones:
\be
\left\{ \begin{array}{ccl}
(011110) &=& 0  \\
(111110) &=& 10\cdot (01102)
\end{array} \right.
\label{k=44}
\ee
i.e. exactly (\ref{k=4}).

\paragraph{The general proof.}
These examples provide an insight for the ansatz for the general recursion. At the generic step $k\longrightarrow k+1 $, one should demonstrate the implication
\be
\left\{ \begin{array}{cclccc}
(\underline{01^j0}) &=& 0
\\
(\underline{1^{j+1}0})  &=&
\frac{(j+1)j}{2}\cdot (01^{j-1}02)
\end{array}\right.
\ \ \ \ \text{ for all } 0\leq j \leq k \ \ \Longrightarrow
\label{jleqk}
\ee
\be
\ \Longrightarrow \
\left\{   \begin{array}{cclcccl}
(\underline{\underline{01^{k+1}0}}) &= &
\frac{(k+1)k}{2}\cdot (001^{k-1}02) &=& ? &=& 0
\\
(\underline{\underline{1^{k+2}0}})  &= &
\frac{(k+1)k}{2}\cdot (1 01^{k-1}02) &=& ? &=&   \frac{(k+2)(k+1)}{2}\cdot (01^{k}02)
\end{array}\right.
\nn
\ee

\noindent
and the proof relies on a pair of substitutions into (\ref{Arel}):
\be
\left\{
\begin{array}{ccc|cccccccccc}
A=[0,2]& k-1 & B=(1^{k-1}0) & \ \ & ( [0,2] \underline{01^{k-1}0 }) &=& ((1^{k-1}0)002)-(0(1^{k-1}0)02)
&=&
\\ &&& \\ \hline &&& \\
A=2 & k & B=(1^k0) & &  (2\underline{01^k0}) &=& ((1^k0)02)-(0(1^k0)2)
&=&
\end{array}
\right.
\nn
\ee

{\footnotesize
\be
\!\!\!\!\!\!\!\!
\left\{
\begin{array}{cccccccccc}
&&& \!\!\!\!\!\!\!\!\!\!\!\!\!\!\!\!\!\!\!\!\!\!\!\!\!\!\!\!\!\!\!\!\!\!\!\!\!\!\!\!\!\!\!\!\!\!\!\!\!\!\!\!\!\!\!\!\!\!\!\!
\delta_{j<k-1}\cdot (1^{k-1-j}\underline{01^{j+2}0}) +\delta_{j,k-1}\cdot (\underline{\underline{01^{k+1}0}})
&- \frac{2}{(j+2)(j+1)}\cdot (\underline{\underline{01^{k+1}0}})&&&&\\
&&&\ \ \ \ \ \ \ S_{001}\ \uparrow &\ \ \ \ \ \ \ (\ref{jleqk})\ \uparrow& &&& \\
&=& \sum_{j=0}^{k-1} (-)^j \frac{(k-1)!}{j!(k-1-j)!}\Big\{\!\!\!\!\!\!\!\!&(1^{k-1-j}01^j\overbrace{002})
&-  (01^{k-1-j}\overbrace{01^j 02}) & \!\!\!\!\!\!\!\!\!\!\!\!\!\!\!\!\!\!\!\!\!\!\!\!\!\!\!\!\!\!\Big\}
\\ &&& \\ \hline &&& \\
&=&  \sum_{j=0}^{k} (-)^j \frac{k!}{j!(k-j)!}\Big\{\!\!\!\!\!\!\!\!&(1^{k-j}\underbrace{01^j02})
&- \delta_{j,0} (01^{k}02)
&\!\!\!\!\!\!\!\!\!\!\!\!\!\!\!- \delta_{j,1}(01^{k-1}\underbrace{012})
&-\delta_{j\geq 2} (01^{k-j}01^{j-2}\underbrace{112})&\Big\} \\
&&&(\ref{jleqk})\ \downarrow  & \downarrow {\rm id}  &  S_{011}\ \downarrow
&\ \ \ \ \ \ \ \ \ \ \ \ \ \ \ \ \ S_{111}\ \downarrow&& \\
&&&\!\!\!\!\!\!\!\!\!\!\!\!\!\!\!\!\!\!\!\!\!\!\!\!\!\!\!\!\!\!\!\!\!\!\!\!\!\!\!\!\!\!\!\!\!\!\!\!\!\!
\delta_{j<k}\cdot\frac{2}{(j+2)(j+1)}\cdot  (\underline{\underline{1^{k+2}0}})+\delta_{j,k}\cdot (01^k02)
& - \delta_{j,0} (01^{k}02) &+ \delta_{j,1}(01^{k}02) &\ \ \ \ \ \ \ \ \ \ \ \ \ \ \ \ \ \ \ \ \ \ \ \ 0&&
\end{array}
\right.
\nn
\ee
}
where we used higher Jacobi identities \cite{HJI}.

\noindent
To summarize, we obtain (\ref{jleqk}):
\be
\left\{
\begin{array}{ccc}
(01^{k+1}0)\cdot \left(1- 2\sum_{j=0}^{k-1}  (-)^j \frac{(k-1)!}{(j+2)!(k-j-1)!}\right) = 0
 \\
\sum_{j}^{k} (-)^j \frac{k!}{j!(k-j)!} \Big\{(1-\delta_{j,k})\cdot\frac{2}{(j+2)(j+1)}\cdot  (1^{k+2}0)
+(\delta_{j,k} -\delta_{j,0}+\delta_{j,1}) (01^k02)\Big\} = 0
\end{array}\right.
\  \Longrightarrow
\nn
\ee
\be
\Longrightarrow \
\left\{
\begin{array}{ccl}
\frac{k-1}{k+1}\cdot(01^{k+1}0)  &=& 0
\\
 (1^{k+2}0)&=&  \frac{(k+2)(k+1)}{2}\cdot (01^k02)
\end{array}
\right.
\ee
The proof of (\ref{intcom}) is completed.

\end{itemize}

\subsection{Properties of the integer rays}

A remarkable fact about this proof {\bf for all integer rays} is that it uses only four lowest Serre relations
$S_{000}$, $S_{001}$, $S_{011}$ and $S_{111}$, still {\it more} than just $S_{000}$.
This means that, despite we are interested in the identity with $E_0$ and $E_1$ only,
$E_2$ also appears in the proof (via $S_{001}$ etc), but {\it not} any higher operator $E_{n\geq 3}$.
A similar observation was already made in \cite{AS} in discussion of the commutativity of time variables $p_k$.
Higher operators, however, contribute to consideration for rational rays in the next section.
Actually, at a given $q$, all operators  $E_{n\leq 2q}$ appear.

Another side of the problem is that the Serre relations appear in many places,
and the actual reduction of $(01^k0)=0$ to them is rather complicated,
as we saw in the simplest example (\ref{01110}).
Generalization of (\ref{01110}) is increasingly complicated, and the above proof is a reasonable substitute.

The third remark is that different integer rays are related by a kind of rotation transformations, but the rotation
in rays the $E$ quarter plane and those in the $F$ quarter plane have to be done in the opposite directions, i.e. by operators inverse to each other. More concretely, there is an operator $\hat O$ such that
\be
\hat E_{i+1}=\hat O\hat E_i\hat O^{-1}\nn\\
\hat F_{i+1}=\hat O^{-1}\hat F_i\hat O
\ee
This operator rotates entire rays of commuting Hamiltonians
and allows one to convert one pair of complementary rays into another, e.g.
\be
H_{\pm k}^{(m)}=\hat O^{\pm m}H_{\pm k}^{(0)}\hat O^{\mp m}
\ee
It was explicitly constructed in \cite[Sec.13]{MMMP} and in terms of that paper it looks like
\be\label{O}
\hat O=\exp\left(-\gamma\hat b_1+ \sum_{m=2}^{\infty} \dfrac{(-1)^{m} \zeta(m)}{m}\,\hat b_m \right)
\ee
where $\zeta(m)$ is the Riemann $\zeta$-function,  $\gamma$ is the Euler constant,
and the commutative family $\{\hat b_m\}$ is related to $\{\hat\Psi_n\}$ by the following
linear transform\footnote{$b_k$ is associated with $W_0(\hat D^k)$ of \cite[Sec.3.4]{MMMP}.}:
\be
\hat\Psi_n: = \sum_{k=0}^{n-1}\binom{n}{k}\,\hat b_k
\ee
However, with this linear transform, (\ref{O}) becomes a somewhat controversial formula in terms of $\{\hat\Psi_n\}$,
and we leave its detailed discussion for another occasion.
The transformation becomes an automorphism in the case of DIM algebra,
when the operators occupy the entire plane.
Then the formulas acquire a more elegant form.

\subsection{Cones}

As it was checked in various representations in \cite{MMMP},
the Hamiltonians associated cones are also commuting. The cone family is given by a set of arbitrary constants $\{\alpha_i\}$ with
the generating operators
\be
\hat E^{(G)}_m=\sum_{i=1}^m\alpha_i\hat E_{i}^{(G)}
\ee
and
\be
\hat E^{(G)}_{m+1}=\sum_{i=1}^m\alpha_i\hat E_{i+1}^{(G)}
\ee
so that the commutative family is given by the Hamiltonians
\be
H^{(G,m)}_k={\rm ad}_{\hat E_{m+1}^{(G)}}^{k-1}\hat E_m^{(G)}
\ee

One can check that the commutativity of these families also follows from the commutativity of the Serre relations.
For instance, consider the generating operator
\be
\hat F^{(G)}_+=\hat E_{m+1}+\alpha\hat E_{m+2}
\ee
Then, the first two Hamiltonians are
\be
\hat H_1=\hat E_{m}+\alpha\hat E_{m+1}\nn\\
\hat H_2=[\hat F^{(G)}_+,\hat H_1]
\ee
They, indeed, commute:
\be
[\hat H_1,\hat H_2]=[\hat E_m,[\hat E_m,\hat E_{m+1}]]+\alpha\left([\hat E_{m+1},[\hat E_m,\hat E_{m+1}]]+
[\hat E_m,[\hat E_{m+1},\hat E_{m+1}]]+[\hat E_m,[\hat E_m,\hat E_{m+2}]]\right)+\nn\\
\alpha^2\left([\hat E_m,[\hat E_{m+1},\hat E_{m+2}]]+
[\hat E_{m+1},[\hat E_m,\hat E_{m+2}]]+[\hat E_{m+1},[\hat E_{m+1},\hat E_{m+1}]]\right)+
\alpha^3[\hat E_{m+1},[\hat E_{m+1},\hat E_{m+2}]]
\ee
The $\alpha^0$ and $\alpha^3$ contributions vanish because they are $S_{000}^{(m)}$ and $S_{000}^{(m+1)}$ correspondingly, the $\alpha$ one, because it is $S_{001}^{(m)}$, and the $\alpha^2$ one, because it is $S_{011}^{(m)}$.

This consideration can be easily extended to arbitrary combinations.

\section{An example for the first rational ray
  \label{sec:attemps-rational-rays}
}

For the rational rays, a reduction of the commutativity to the Serre relations does not work,
at least in a straightforward way.
Hence, our conclusion is that the commutativity for the rational rays
uses also the quadratic relations \eqref{eq:serre-quadratic},
and is not preserved by the $\beta$-deformation (unless changing the very definition of the rational ray).
Since these are a kind of {\it negative} statements,
it is sufficient to present a detailed example in the simplest case.

\subsection{A possible notation}

As we already discussed around formula (\ref{calH2}),
for the family ${ H}^{(2m+1,2)}_k$ we need to prove a sequence of identities
\be
(\tilde 0\tilde 1^{k}\tilde 0)=0
\label{2ratcom}
\ee
with
\be
\tilde 0 := [\hat E_0,\hat E_2] \ \ \ {\rm and} \ \ \ \ \tilde 1 :=[\hat E_1,\hat E_2] +  [\hat E_0,\hat E_3]
\label{Ham22}
\ee
instead of $0:=\hat E_0$ and $1:=\hat E_1$ from the previous section.
This notation makes (\ref{2ratcom}) into direct analogues of (\ref{intcom}),
however, the generalization of the proof is not at all immediate.
One may also think that, behind this generalization, there
is a kind of rotation of the algebra generators
\be
\hat E_i \longrightarrow e^{-t\hat E_n}\hat E_ie^{t\hat E_n}
= \hat E_i + t\cdot [\hat E_i,\hat E_n] + \ldots
\label{2ratrot}
\ee
modified appropriately to preserve the Serre relations.
Let us see to what extent this analogy can work.

If (\ref{2ratrot}) is to be taken seriously, the simplest version $(\tilde 0\tilde 0\tilde 1)=0$ of (\ref{2ratcom}),
the deformed Serre relation $S_{001}=(001)$
would appear in the cubic order in $t$.
It makes sense first to look at the first order:
\be
(\tilde 0 0 1) + (0 \tilde 0 1) + (00\tilde 1)
= ([0,2]01) + (0[0,2]1) + (0012) + (0003) =
\nn \\
= (0\underbrace{201})-(2\underbrace{001}) -(0\underbrace{102}) + (0\underbrace{012})+(0\underbrace{003})
= 2\cdot (0012)-(0102) \ \stackrel{S_{011}}{=} \ 3\cdot(0012)
\ee

\noindent
The first and the last terms combine in the Serre relation $\frac{1}{2}S_{002} = (003)+(201)-(012) = 0$,
the second term vanishes due to $\frac{1}{6}S_{000} = (001)=0$,
but the net result is not the Serre relation and does not vanish.
Still some cancellations occur, and this can cause cautious optimism.
Perhaps, most important is that we see a reason for an additional term $[0,3]$ in $\tilde 1$,
which is beyond the naive suggestion (\ref{2ratrot}).

\subsection{Serre is not enough\label{sec:attempts-rational-rays}}

We could wish to prove that $(\tilde 0\tilde 0 \tilde 1)$ is a combination of Serre relations,
but we fail. This is not surprising: as we demonstrated in \cite{MMMP}, it does not vanish upon the $\beta$-deformation, though the Serre relations are still correct after the deformation. We demonstrated this in concrete representations, however, as soon as we are making here statements at the level of algebra, it is enough to show it is wrong in a concrete representation.
Still, it deserves reviewing the attempt.

First of all, this quantity is a linear combination of items, each containing exactly six $E_i$
with the total grading $i_1+\ldots+i_6=7$.
This means that no quantities with $E_i$, $i>7$ can contribute.

Second, $(\tilde 0\tilde 0 \tilde 1)$ per se contains at most $E_3$, and only in the combinations $000223$.
The only Serre relations which do not contain $1$, are $S_{222}$, $S_{022}$ and $S_{002}$.
However  it turns out to be {\it impossible} to eliminate all the terms with $E_3$ from  $(\tilde 0\tilde 0 \tilde 1)$
with just the help of them.
This may seem to be a problem, but from the experience in the previous section (with integer rays)
we already know what to do.

A way out is to consider the Serre relations with $E_4$ in peculiar combinations,
where $E_4$ is canceled.
There are four combinations with this property:
\be
(411S_{000})-2\cdot(141S_{000})+(114S_{000}) + 3\cdot(100S_{113})-6\cdot(010S_{113})+3\cdot(001S_{113})
\label{E41}
\ee
\be
(420S_{000}) + (402S_{000})-2\cdot(240S_{000}) -2\cdot (042S_{000})+(204S_{000})+(024S_{000})
+ \nn \\
+6\cdot (100S_{023})-12\cdot (010S_{023}) + 6\cdot(001S_{023})
\ee
\be
(410S_{001})+(401S_{001}-2\cdot(140S_{001}) +(104S_{001})-2\cdot(041S_{001})+(014S_{001})
+ \nn \\
+ 2\cdot (200S_{013})-4\cdot (020S_{013}) +2\cdot(002S_{013})
\ \ \ -2\cdot(110S_{013})+4\cdot(101S_{013})-2\cdot (011S_{013})
\ee
and
\be
(210S_{003}) +(201S_{003})-\!2\!\cdot\! (120S_{003})-\!2\!\cdot\! (021S_{003})+(102S_{003})+(012S_{003})
 +\nn \\
+(400S_{011})-\!2\!\cdot\!(040S_{011})+(004S_{011})
\label{E44}
\ee
They do not contain $E_4$, and
in the $E_3$ sector they contain the sets like $001123$ in addition to $000223$.
This allows one to get more Serre relations involved,
from  which we can also get new terms with $000223$ to potentially resolve our problems.
$E_4$ may appear only once, to preserve the selection rule $i_1+\ldots + i_6=7$.

Unfortunately, this time adding items with $E_4$ does not help.
We can attempt to add also $E_5$, $E_6$ and $E_7$ in combinations, where they all cancel,
together with $E_4$.
We can not add more, because of the grading constraint $i_1+\ldots+i_6=7$.
Of course, adding higher $E_i$ brings new terms with $E_4$, and
modifies the structures (\ref{E41})-(\ref{E44}).

In fact, $E_7$ could appear only from $S_{006}$, and there is no way to cancel it afterwards.
Thus the highest real option is with $E_6$
There is just a single combination, where all items contain $E_6$, but the sum does not:
\be
 (600S_{000}) - 2\cdot (060S_{000}) +  (006S_{000})
+3 \cdot (100S_{005})- 6\cdot (010S_{005})+3\cdot (001S_{005})
\ee
Note that the structure $(000S_{015})$  also has the proper grading,
but it does not enter this combination.

Still, the terms with $E_7,E_6,E_5,E_4$ do not help:
$(\tilde 0 \tilde 1^k \tilde 0)$ is {\it not} a combination of the Serre identities,
already for $k=1$.

\subsection{$k=1$ beyond Serre}

However, if we allow one to use the quadratic relations \eqref{eq:serre-quadratic} along with the Serre ones,
the problem disappears.
Actually, we need only the second relation from the list, which we denote by
\be
R_{jk}:=[\hat E_{j+3},\hat E_k]-3[\hat E_{j+2},\hat E_{k+1}]+3[\hat E_{j+1},\hat E_{k+2}]-[\hat E_{j},\hat E_{k+3}]
+[\hat E_{j+1},\hat E_{k}]-[\hat E_{j},\hat E_{k+1}]=0
\ee
Then
\begin{align}
(\tilde 0 \tilde 1 \tilde 0) = & \
\sum_{i_1,i_2,i_3,i_4=0}^2 \delta_{i_1+i_2+i_3+i_4,4}\cdot v_{i_1i_2i_3i_4}\cdot  (i_1,i_2,i_3,i_4,R_{00})
+ \nn \\
+ & \ \sum_{i_1,i_2,i_3=0}^3 \ \ \sum_{0\leq j_1\leq j_2\leq j_3\leq 2}
\left ( \delta_{i_1+i_2+i_3+j_1+j_2+j_3,6} + \delta_{i_1+i_2+i_3+j_1+j_2+j_3,4} \right )
\cdot
u_{i_1i_2i_3, j_1j_2j_3}\cdot
(i_1,i_2,i_3,S_{j_1j_2j_k})
\end{align}
The delta-symbols account for the gradings and reduce the number of possible terms in the sums,
and there are two delta-symbols in the last summand because $R_{00}$
is not homogeneous in the sum of $\hat{E}$'s indices.
It is sufficient to use only $R_{00}$, but it enters with non-vanishing coefficients.
A convenient symmetric choice is
\be
v_{1111} = -\frac{3}{2}, \nn \\
v_{2200}=v_{2020} = v_{2002} = v_{0220} = v_{0202} = v_{0022} = -\frac{1}{2}, \nn \\
\underbrace{v_{2110} = \ldots =v_{0112}}_{12\ {\rm structures}} = -\frac{3}{4}\
\ee
After that, there is a big freedom in choosing the coefficients $u$, i.e. a combination of the Serre
relations.
Note that the highest appearing operator is $\hat E_3$.
In contrast with $v$, we cannot choose $u$ to be symmetric w.r.t its indices,
not even w.r.t. its first three indices.\footnote{
A sample explicit solution  is
\begin{align}\scriptscriptstyle
u_{0, 1, 0, 1, 1, 1} = & \ \scriptscriptstyle 1/12, u_{0, 1, 1, 0, 1, 1} = -1/4,
 u_{0, 1, 1, 0, 2, 2} = -4/9, u_{0, 1, 1, 1, 1, 2} = 2/9,
 u_{0, 1, 2, 0, 0, 1} = -1/2, u_{0, 2, 0, 0, 1, 1} = 1/2,
 u_{0, 2, 0, 0, 2, 2} = -2/3, u_{0, 2, 0, 1, 1, 2} = 10/9,
 \notag \\ \notag \scriptscriptstyle
 u_{0, 2, 1, 0, 0, 1} = & \ \scriptscriptstyle -1, u_{0, 2, 1, 0, 1, 2} = -40/9,
 u_{0, 2, 1, 1, 1, 1} = 106/27, u_{0, 2, 2, 0, 0, 0} = -1/2,
 u_{0, 2, 2, 0, 0, 2} = -2/3, u_{0, 2, 2, 0, 1, 1} = 26/9,
 u_{0, 2, 3, 0, 0, 1} = -2/3,
 \\ \notag \scriptscriptstyle
 u_{0, 3, 0, 1, 1, 1} = & \ \scriptscriptstyle 13/108,
 u_{0, 3, 1, 0, 1, 1} = -53/36, u_{0, 3, 2, 0, 0, 1} = -1/2,
 u_{1, 0, 0, 1, 1, 1} = 3/4, u_{1, 0, 0, 1, 2, 2} = 7/30,
 u_{1, 0, 1, 0, 1, 1} = -11/4, u_{1, 0, 1, 0, 2, 2} = -2/45,
 \\ \notag \scriptscriptstyle
 u_{1, 0, 1, 1, 1, 2} = & \ \scriptscriptstyle -73/90, u_{1, 0, 2, 0, 0, 1} = 1/2,
 u_{1, 0, 2, 1, 1, 1} = -46/45, u_{1, 0, 3, 0, 1, 1} = 8/5,
 u_{1, 1, 0, 0, 1, 1} = 5/2, u_{1, 1, 0, 0, 2, 2} = -19/90,
 u_{1, 1, 0, 1, 1, 2} = 4/3,
 \\ \notag \scriptscriptstyle
 u_{1, 1, 1, 0, 0, 1} = & \ \scriptscriptstyle -3/2,
 u_{1, 1, 1, 0, 1, 2} = -101/45, u_{1, 1, 1, 1, 1, 1} = 152/135,
 u_{1, 1, 2, 0, 0, 0} = 1/2, u_{1, 1, 2, 0, 0, 2} = -11/6,
 u_{1, 1, 2, 0, 1, 1} = -409/90, u_{1, 1, 3, 0, 0, 1} = -7/30,
 \\ \notag \scriptscriptstyle
 u_{1, 2, 0, 0, 1, 2} = & \ \scriptscriptstyle -107/45, u_{1, 2, 0, 1, 1, 1} = 1897/540,
 u_{1, 2, 1, 0, 0, 0} = -11/6, u_{1, 2, 1, 0, 0, 2} = -77/90,
 u_{1, 2, 1, 0, 1, 1} = 163/12, u_{1, 2, 2, 0, 0, 1} = 121/45,
 \\ \notag \scriptscriptstyle
 u_{1, 2, 3, 0, 0, 0} = & \ \scriptscriptstyle -13/54, u_{1, 3, 0, 0, 1, 1} = -79/45,
 u_{1, 3, 1, 0, 0, 1} = 7/90, u_{1, 3, 2, 0, 0, 0} = -7/54,
 u_{2, 0, 0, 0, 1, 1} = 3/10, u_{2, 0, 0, 0, 2, 2} = 13/75,
 u_{2, 0, 0, 1, 1, 2} = -19/450,
 \\ \notag \scriptscriptstyle
 u_{2, 0, 1, 0, 0, 1} = & \ \scriptscriptstyle -1,
 u_{2, 0, 1, 0, 1, 2} = 38/225, u_{2, 0, 1, 1, 1, 1} = -683/675,
 u_{2, 0, 2, 0, 0, 0} = 7/30, u_{2, 0, 2, 0, 0, 2} = -13/15,
 u_{2, 0, 2, 0, 1, 1} = 176/225, u_{2, 0, 3, 0, 0, 1} = 52/75,
 \\ \notag \scriptscriptstyle
 u_{2, 1, 0, 0, 0, 1} = & \ \scriptscriptstyle 2/5, u_{2, 1, 0, 0, 1, 2} = 287/225,
 u_{2, 1, 0, 1, 1, 1} = -3733/2700, u_{2, 1, 1, 0, 0, 0} = -7/15,
 u_{2, 1, 1, 0, 0, 2} = -481/450, u_{2, 1, 1, 0, 1, 1} = -1487/300,
 \\ \notag \scriptscriptstyle
 u_{2, 1, 2, 0, 0, 1} = & \ \scriptscriptstyle -1997/450, u_{2, 1, 3, 0, 0, 0} = 287/1350,
 u_{2, 2, 0, 0, 0, 0} = -7/15, u_{2, 2, 0, 0, 0, 2} = -12/25,
 u_{2, 2, 0, 0, 1, 1} = -1/75, u_{2, 2, 1, 0, 0, 1} = 191/75,
 \\ \notag \scriptscriptstyle
 u_{2, 2, 2, 0, 0, 0} = & \ \scriptscriptstyle 37/450, u_{2, 3, 0, 0, 0, 1} = -13/25,
 u_{2, 3, 1, 0, 0, 0} = -259/1350, u_{3, 0, 0, 1, 1, 1} = -5/108,
 u_{3, 0, 1, 0, 1, 1} = 37/36, u_{3, 0, 2, 0, 0, 1} = 1/2,
 u_{3, 1, 0, 0, 1, 1} = 4/9,
 \\ \notag \scriptscriptstyle
 u_{3, 1, 1, 0, 0, 1} = & \ \scriptscriptstyle -2/9,
 u_{3, 1, 2, 0, 0, 0} = 4/27, u_{3, 2, 1, 0, 0, 0} = -2/27
 \nn
\end{align}
This solution is not unique and is not adjusted to be the simplest one in any sense,
we present it just as a random example.
}

\subsection{Another $H^{[12]}_2$}

Since $R_{00}=0$ expresses $[\hat E_3,\hat E_0]$ through $[\hat E_2,\hat E_1]$, one can also modify (\ref{Ham22})
\be
\tilde 0 := [\hat E_0,\hat E_2] \ \ \ {\rm and} \ \ \ \ \tilde{\tilde 1} :=[\hat E_1,\hat E_2]
\label{Ham220}
\ee
to get slightly simplified expressions:
\begin{align}
(\tilde 0 \tilde{\tilde 1} \tilde 0) = & \
\sum_{i_1,i_2,i_3,i_4=0}^2 \delta_{i_1+i_2+i_3+i_4,4}\cdot \tilde v_{i_1i_2i_3i_4}\cdot  (i_1,i_2,i_3,i_4,R_{00})
+  \\
+ & \sum_{i_1,i_2,i_3=0}^3 \ \ \sum_{0\leq j_1\leq j_2\leq j_3\leq 2}
\left ( \delta_{i_1+i_2+i_3+j_1+j_2+j_3,6}
+ \delta_{i_1+i_2+i_3+j_1+j_2+j_3,4} \right )\cdot
\tilde u_{i_1i_2i_3, j_1j_2j_3}\cdot
(i_1,i_2,i_3,S_{j_1j_2j_k})  \nn
\end{align}
with
\be
\tilde v_{1111} = \frac{1}{3}, \nn \\
\tilde v_{2200}=\tilde v_{2020} =\tilde  v_{2002} =\tilde  v_{0220} = \tilde v_{0202} = \tilde v_{0022} = \frac{1}{6}, \nn \\
\underbrace{\tilde v_{2110} = \ldots =\tilde v_{0112}}_{12\ {\rm structures}} = \frac{1}{6}\
\ee
and with the corresponding $\tilde u$, which can again be chosen in many different ways.

\section{Summary}

In summary, our attempt in this section demonstrates that the Serre relations
\eqref{eq:serre-cubic} are not enough to show commutativity
even of the first two Hamiltonians and even on the first essentially rational ray $(1,2)$.
Commutativity is obtained only if one takes into account the
quadratic relations \eqref{eq:serre-quadratic} as well.
This means that commutativity of rational ray Hamiltonians
after the $\beta$-, and $(q,t)$-deformations is not \textit{a priori} guaranteed,
and some \textit{fine-tuning} is required. However, commutative rational ray families can be ultimately
constructed in these cases too, since the $(q,t)$-deformation of the doubled $W_{1+\infty}$ algebra is the
Ding-Iohara-Miki (DIM) algebra \cite{DI,Miki}, and, in this latter, there are Heisenberg subalgebras associated with each rational
line \cite{Smirnov}. This follows from the $SL(2,\mathbb{Z})$-automorphism of the Ding-Iohara-Miki algebra \cite{Miki1,Miki}.
The commutative half of the Heisenberg subalgebra is just the commutative family of Hamiltonians associated with the rational ray. However, the limiting procedure to the affine Yangian is not that immediate \cite{Ch3}.

Note that, in the DIM algebra case, one can also start with the Lie algebra $\overline{W_{1+\infty}}$ \cite{Miki},\cite[Appendix A1]{A7}, which is a double of the $W_{1+\infty}$ algebra  (hence, Heisenberg subalgebras instead of commutative families) and may have two central extensions, and reformulate it in terms of a few generating elements subject to commutation relations and Serre relations. In this case, the Serre relations looks similar to (\ref{eq:serre-cubic}). One can further deform this algebra to the DIM algebra by deforming all relations but the Serre ones. Hence, the reasoning of this paper is literally applicable to the DIM algebra: one can generate the Heisenberg subalgebras associated with the integer rays basing solely on the Serre relations exactly in the same way as in sec.3: by similar commutators, while the Heisenberg subalgebras associated with the rational rays need some additional treatment.

\section{Conclusion}

The goal of this paper was to reduce the discovery of non-trivial systems of commuting Hamiltonians
\cite{MM,MMMP}
to the properties of $W_{1+\infty}$ algebra and, more generally, of the affine Yangian of $\mathfrak{gl}_1$.
This explains why commutativity holds in a big variety of representations
as was demonstrated explicitly in \cite{MMMP}.

The main result is the proof that this works perfectly for {\bf integer rays},
where the  relevant relations are just the Serre relations,
which are independent of the
parameters $\sigma_2=h_1h_2+h_2h_3+h_3h_1$ and $\sigma_3=h_1h_2h_3$
describing the deformation of the $W_{1+\infty}$ algebra to the affine Yangian, and hence correct for the both.
This explains why the commutativity is preserved after the $\beta$-deformation
with $\sigma_2=-1-\beta(\beta-1)$ and $\sigma_3=-\beta (\beta-1)$,
which was empirically  observed in \cite{MMMP}.
This also implies commutativity for other representations not discussed in \cite{MMMP} like the MacMahon representation \cite{Prochazka} or a still hypothetical ``triangular-time" representation of \cite{MTs}.

On the contrary, we did not succeed in getting the same result for more general {\bf rational rays}.
The commutativity in this case depends also on the quadratic relations in algebra,
which are substantially deformed for $\beta\neq 1$.
This is again in accordance with the empirical observation of \cite{MMMP},
where commutativity along the rational rays was violated by the $\beta$-deformation,
both in time and in eigenvalue representations.
We considered just a single example of the simplest pair of rational Hamiltonians
$[H^{2,1}_2,H^{2,1}_1] \sim (\beta-1)$,
leaving the general proof for further publications.

The case of {\it cones} (generalization of rays outlined in \cite{MMMP})
celebrates the same commutativity properties inherited
from rays, i.e. commutativity is preserved by $\beta$-deformation of conic combinations
of integer rays, and is violated if admixture of rational rays is present.
A separate issue is the case of ``vertical cones", where the commuting Hamiltonians also exist
but are just associated with the Cartan subalgebra of the both $W_{1+\infty}$ algebra and the affine Yangian.

\section*{Acknowledgements}

This work was supported by the Russian Science Foundation (Grant No.21-12-00400)

\end{document}